%% file: conf_paper.tex
\documentclass[11pt]{article}
\usepackage{graphicx}
\usepackage{subfigure}

\newcommand{\BABARPubYear}    {03}

\newcommand{\BABARConfNumber} {005}
\newcommand{\SLACPubNumber} {9688}

\input{babarsym}

\long\def\inst#1{\par\nobreak\kern 4pt\nobreak
    {\it #1}\par\vskip 10pt plus 3pt minus 3pt}

\input{user_symbols}

\begin{document}
{\pagestyle{empty}

\begin{flushright}
\babar-CONF-\BABARPubYear/\BABARConfNumber \\
SLAC-PUB-\SLACPubNumber \\
March 2003 \\
\end{flushright}

\par\vskip 5cm

\begin{center}
\Large \bf A Search for \mbox{\boldmath$\Bu \to \taup \nu_{\tau} $\unboldmath} Recoiling Against \mbox{\boldmath$\btdlnux$\unboldmath}
\end{center}
\bigskip

\begin{center}
\large The \babar\ Collaboration\\
\mbox{ }\\
\today
\end{center}
\bigskip \bigskip

\begin{center}
\large \bf Abstract
\end{center}
We present a search for the decay \btn\ in $88.9 \times 10^{6}$ $\FourS$
decays recorded with the \babar\ detector at the SLAC $B$-Factory. A
sample of semi-exclusive, semi-leptonic $B$ decays (\btdlnux), where $X$
is either a photon, \piz, or nothing, is used to identify the daughter
particles that are associated with the other $B$ meson in each event.
These particles are searched for evidence of a one-prong leptonic \btn\
decay. For this data sample we set a preliminary upper limit on the branching
fraction of $\mathcal{B}(\btn) < 4.9 \times 10^{-4}$ at the 90\%
confidence level. This result is then combined with a statistically
independent \babar\ search for \btn\ to give a combined preliminary limit of
$\mathcal{B}(\btn) < 4.1 \times 10^{-4}$.

\vfill
\begin{center}
Presented at the XXXVIII$^{th}$ Rencontres de Moriond on\\
Electroweak Interactions and Unified Theories, \\
3/15---3/22/2003, Les Arcs, Savoie, France
\end{center}

\vspace{1.0cm}
\begin{center}
{\em Stanford Linear Accelerator Center, Stanford University, 
Stanford, CA 94309} \\ \vspace{0.1cm}\hrule\vspace{0.1cm}
Work supported in part by Department of Energy contract DE-AC03-76SF00515.
\end{center}

\newpage
} 

\input{authors_win2003.tex}

\setcounter{footnote}{0}

\section{Introduction}
\label{sec:Introduction}
\begin{figure}[htb]
\begin{center}
\includegraphics[width=0.45\textwidth]{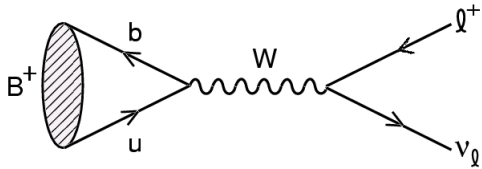}
\end{center}
\caption{\label{fig:feynman_diagram}%
Purely leptonic $B$ decay $\Bu \to \ell^{+} \nu_{\ell}$  
proceeds via quark annihilation into a $W$ boson.}
\end{figure}

In the Standard Model, the purely leptonic decay \btn\ 
\footnote{Charge-conjugate modes are implied throughout this paper.%
The signal $B$ will always be denoted as a \Bu\ decay while the
semi-leptonic $B$ will be denoted as a \Bub\ to avoid confusion.}
proceeds via quark annihilation 
into a $W$ boson (Fig. \ref{fig:feynman_diagram}).
Its amplitude is thus proportional to the product of the 
$B$-decay constant, $f_B$ and the quark-mixing-matrix 
element $V_{ub}$. The branching fraction is given by:
\begin{equation}
\label{eqn:br}
\mathcal{B}(B^{+} \rightarrow {\taup} \nu)= 
\frac{G_{F}^{2} m^{}_{B}  m_{\tau}^{2}}{8\pi}
\left[1 - \frac{m_{\tau}^{2}}{m_{B}^{2}}\right]^{2} 
\tau_{\Bu} f_{B}^{2} \mid V_{ub} \mid^{2},\label{eq:brfrac} 
\end{equation}
where we have set $\hbar = c = 1$, %
$G_F$ is the Fermi constant, 
$V_{ub}$ is the quark mixing matrix element \cite{ref:c} \cite{ref:km}, 
$f_{B}$ is the $\Bu$ meson decay constant, describing the
overlap of the quark wave-functions inside the meson,
$\tau_{\Bu}$ is the $\Bu$ lifetime and
$m^{}_{B}$ and $m_{\tau}$ are the $\Bu$ meson and $\tau$ masses.
This expression is entirely analogous to that for pion decay.
Physics beyond the Standard Model, such as SUSY, could
enhance this process, as through the introduction of a charged Higgs 
\cite{ref:mssm}.

Current theoretical values for $f_B$ have  
large uncertainty, and purely leptonic decays of the $\Bu$ meson
may be the only clean experimental method of measuring
$f_B$ precisely. Given measurements of $V_{ub}$ from semi-leptonic
processes such as $B \to \pi \ell \nu$, $f_{B}$ could be extracted
from the measurement of the \btn\ branching fraction. In addition,
by combining the branching fraction measurement with results
from $B$ mixing the ratio $|V_{ub}|/|V_{td}|$ can be extracted
from $\mathcal{B}(\btn)/\Delta m$, where $\Delta m$ is the
mass difference between the heavy and light neutral $B$ meson states.

The decay amplitude is proportional to the lepton mass and
decay to the lighter leptons is severely suppressed. This mode
is thus the most promising for discovery at existing experiments.
However, challenges 
such as the large missing momentum from several neutrinos make 
the signature for \btn\ less distinctive than for other leptonic modes.

The Standard Model estimate of this branching fraction is $7.5 \times 10^{-5}$,
given the PDG 2002 values for $f_B$, $|V_{ub}|$, and the $\Bu$ lifetime: 
$f_{B} = 198\mev$, $V_{ub} = 0.0036$,  $\tau_{\Bu} = 1.674\ps$
\cite{ref:pdg2002}. 

Purely leptonic $B$ decays have not yet been observed. CLEO \cite{cleo}  and experiments 
at LEP \cite{l3}\cite{aleph}\cite{delphi} have searched for this process and set limits on the 
branching fraction at the 90\% confidence level (CL).
The most stringent limit on $\mathcal{B}(\btn)$ comes from the
L3 experiment:
\begin{eqnarray}
\mathcal{B}(\btn) & < & 5.7 \times 10^{-4} \, \textrm{at the 90\% CL }
\end{eqnarray}

\section{The \babar\ Detector and Dataset}
\label{sec:babar}
The data used in this analysis were collected with the \babar\ detector
at the \pep2\ storage ring. The sample corresponds to an integrated
luminosity of $81.9\invfb$ at the \FourS\ resonance (on-resonance) and $9.58\invfb$
taken $40\mev$ below the \FourS\ resonance (off-resonance). The on-resonance sample consists of
about 88.9 million \BB\ pairs. The collider is operated with asymmetric
beam energies, producing a boost of $\beta\gamma \approx 0.56$ 
of the \FourS along the collision axis.

The \babar\ detector is described elsewhere~\cite{ref:babar}.
Charged-particle momentum, direction, and energy loss ($\frac{dE}{dx}$) 
measurements are performed
by a five-layer double-sided silicon vertex tracker (SVT) and a 40-layer drift 
chamber (DCH) which are immersed in the field of a 1.5-T super-conducting 
solenoid.
Charged particle identification is performed by a detector of
internally-reflected Cherenkov light (DIRC). The energies of 
electrons and photons are measured by the electromagnetic 
calorimeter (EMC) consisting of
an array of 6580 CsI(Tl) crystals. The flux return is instrumented
with resistive plate chambers (IFR) to detect the passage of
muons and neutral hadrons.

A GEANT4-based \cite{geant4} Monte Carlo (MC) 
simulation is used to model the signal
efficiency and the physics backgrounds. Simulation samples
equivalent to approximately three times the accumulated data  were
used to model \BB\ events, and samples equivalent to approximately
1.5 times the accumulated data were used to model $\epem \to$
\uubar, \ddbar, \ssbar, \ccbar, and \tautau\ events.

\section{Analysis Method}
\label{sec:Analysis}

\subsection{Event Selection}

The decay \btn\, and subsequent one-prong leptonic decays
$\taup \to \ep \nue \nutb$ and $\taup \to \mup \num \nutb$ 
lead to the production of a single, visible charged
particle and missing energy carried by neutrinos.
Therefore, any remaining neutral energy or charged tracks
originate from the other $B$. The search 
for \btn\ proceeds by reconstructing one of the
$B$ mesons in each event in the semi-leptonic topology \btdlnux\
(henceforth referred to as the \emph{semi-leptonic side} or 
\emph{semi-leptonic B}),
where $X$ could be a photon, neutral pion, or nothing. The
remainder of the event is searched for evidence of
the one-prong leptonic \taup decay modes. 

The choice of a semi-exclusive, semi-leptonic $B$ decay means that the $B$ meson
cannot be fully reconstructed, since there is a neutrino missing
and excited neutral $D$ states are not reconstructed, potentially
leaving unassigned neutral energy in the event. The inherent
cleanliness of the events (small multiplicity, little unassigned
neutral energy), however, is a powerful constraint that compensates
for the choice of reconstruction mode. This method has been used in a
search for $\Bu \to K^{+} \nu \bar{\nu}$ \cite{knunu}.

Events are pre-selected by requiring that their net charge is zero,
that $R_{2}<0.9$ (where $R_2$ is the normalized second
Fox-Wolfram moment \cite{foxwolfram}), and that they have large missing mass 
($M_{miss}>1.0\gev$). 
We select our semi-leptonic side by first reconstructing \Dz\ candidates in
one of four modes: $\Dz \to K^{-} \pi^{+}$, $\Dz \to K^{-} \pi^{+} \piz$,
$\Dz \to K^{-} \pi^{+} \pi^{+} \pi^{-}$, and $\Dz \to \KS \pi^+ \pi^-$.
The \KS\ is reconstructed only in the charged mode $\KS \to \pi^{+}\pi^{-}$.
The \Dz\ meson is required to have a momentum in the center-of-mass (\FourS)
frame $p^{*}_{\Dz}>0.5\gev$, and its daughters are required to
meet particle identification criteria consistent with the particle hypothesis.

\begin{figure}[htb]
\begin{center}
\includegraphics[width=0.45\textwidth]{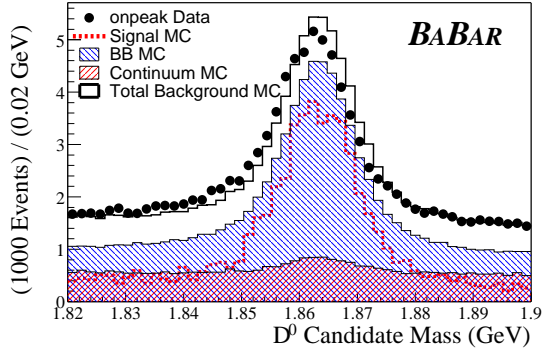}
\end{center}
\caption{\label{fig:d0mass}%
The \Dz\ mass distribution after making event preselections and
requiring that events have at least one \btdlnux\ candidate. 
Background MC combines \BB\ (excluding \btn\ decays) and 
continuum MC and is normalized to the data area, 
while the normalization of the 
signal simulation is arbitrary. Due to the large size of
this sample, the data error bars are too small to see in the figure.}
\end{figure}

We pair the \Dz\ candidates with leptons whose momentum in the
center-of-mass frame meets the requirement $p_{\ell}^{*}>1.3\gev$. 
The $\Dz$-lepton pairs are required to converge at a common
vertex; these pairs are referred to as \Dl\ candidates. If there
are multiple \Dl\ candidates in an event we select the
candidate whose reconstructed \Dz\ mass is closest to the average
reconstructed mass for its decay mode. If the best
candidate can be paired with a pion to make a $\Dstarp\ell^{-}$
candidate, then the event is consistent with being a $\Bz\Bzb$
event and is rejected.

The mass distribution for these $D^{0}$ candidates is shown in
Fig. \ref{fig:d0mass}. There is a difference between the mean
of data and the mean of MC because the MC generator uses a \Dz\ mass
that differs from the PDG value by $1\mev$.
The width of distribution is typically narrower in MC by $0.2\mev$ 
($K^{-} \pi^{+}$) to $4 \mev$ ($K^{-} \pi^{+} \piz$) 
due to energy and momentum resolution differences 
in data and MC. To account for differences in mean
position and peak width, we require that the reconstructed \Dz\ mass be within 
$3\sigma$ of the fitted mean for the mode under consideration. 
The widths of the peaks vary by \Dz\ mode, from $3.8\mev$ for
$\Dz \to K^{-} \pi^{+} \pi^{-} \pi^{+}$ to $10.4\mev$ for 
$\Dz \to K^{-} \pi^{+} \piz$.

\begin{figure}[hbt]
\begin{center}
\includegraphics[width=0.45\textwidth]{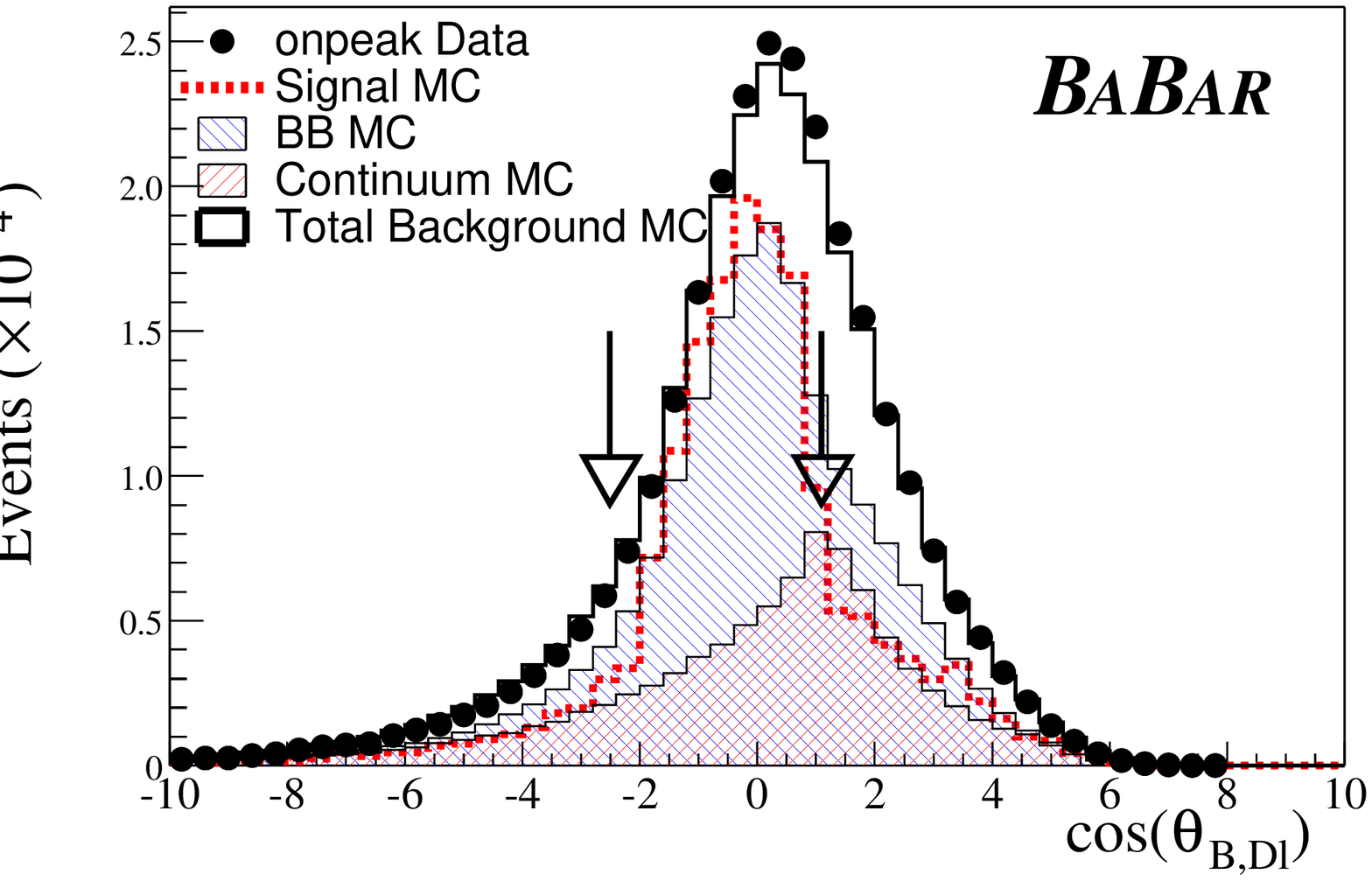}
\end{center}
\caption{\label{fig:cosbydl}%
The quantity \cosbydl, computed from the energies and momenta of the \Dl\ 
and the $B$ meson assuming only a neutrino is missing.
Background MC combines \BB\ (excluding \btn\ decays) and continuum
MC and is normalized to the
data area, while the normalization of the signal simulation is arbitrary.
The arrows
represent the boundaries of the selection region between $-2.5$ and $1.1$. Due to the
large size of this sample, the data error bars are too small to see 
in the figure.
}
\end{figure}

A restriction is also placed on \cosbydl, where $\theta_{B,D\ell}$ is
computed from $B$ and \Dl\ momenta and energy in the center-of-mass frame 
under the assumption that only a neutrino is missing,
\begin{equation}
\cosbydl = \frac{ (2\,E_{{\B}} E_{\Dl} - m^2_{{\B}} - m^2_{\Dl})}
{2\,|\vec{p}_{{\B}}| |\vec{p}_{\Dl}|   }\, .
\end{equation}
Since the semi-leptonic $B$ cannot be completely reconstructed its energy, $E_{{\B}}$,
and momentum, $|\vec{p}_{{\B}}|$,
are derived from the beam energies. The variable \cosbydl\ can take on 
values outside $(-1,1)$ for events in which particles besides the
neutrino have been missed in the semi-leptonic decay.
We allow for the feed-down 
of excited neutral $D$ states into the analysis by
applying an asymmetric cut of $-2.5<\cosbydl<1.1$. This variable is shown in
Fig. \ref{fig:cosbydl}. Differences in the semi-leptonic side between 
data and simulation are later corrected 
(Sec. \ref{sec:Systematics_efficiency}) in the overall 
efficiency estimate.

\begin{figure}[hbt]
\begin{center}
\includegraphics[width=0.45\textwidth]{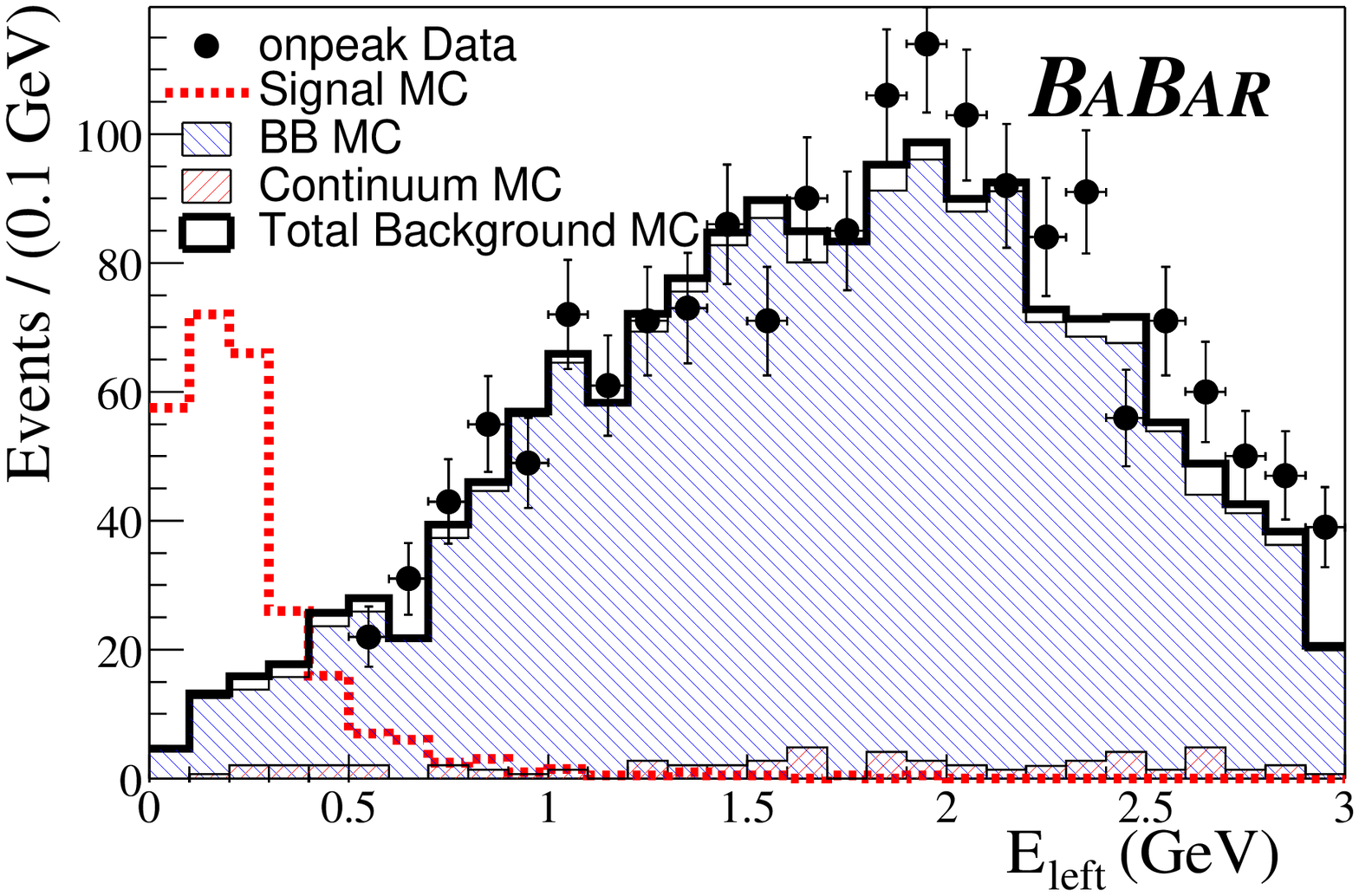}
\end{center}
\caption{\label{fig:eleft}%
\eleft, the neutral energy
remaining in the calorimeter after excluding neutrals associated with the
semi-leptonic side. In the above distribution all analysis selection
criteria are applied. The data are not shown below $0.5\gev$. 
The normalization of the signal MC is arbitrary.}

\end{figure}

Once the semi-leptonic side has been reconstructed, its daughter particles
are excluded from the event. Any neutral or charged object
that is not associated with the semi-leptonic side is referred to as belonging to the
\emph{signal side}. The event is then
required to have only one remaining charged particle with a small
impact parameter in the $(r,\phi)$ plane. 
This track is required to fail kaon identification
and be identified either as a muon or an electron. 
In addition, \tautau\ continuum events
are rejected by making requirements on the angle
of the track with respect to the event thrust axis 
($|\cos{\theta_{\vec{p},\vec{T}}}|<0.9$) and the
minimum mass capable of being made from any three tracks in the
event ($M_{3}^{min}>1.5\gev$). In general, continuum events
are sharply peaked toward 1.0 in $|\cos{\theta_{\vec{p},\vec{T}}}|$;
additionally, \tautau\ events tend to have a value of $M_{3}^{min}$
peaked below the $\tau$ mass.

The variable \eleft,
the neutral energy remaining in the calorimeter
after excluding neutrals from the semi-leptonic $B$, 
provides good discrimination
between signal and background (Fig. \ref{fig:eleft}). 
For true signal events,
this quantity is peaked toward zero energy while for background it peaks
near $2\gev$. We make a loose
restriction of $\eleft<1.0\gev$ in preparation for the final step of the analysis,
which is to fit the \eleft\ distribution and extract the signal and 
background contributions. 

To minimize experimenter bias, the data were not studied in the
signal-populated region of \eleft\ until systematic checks were
performed using control samples in the region of \eleft\ 
($\eleft>0.5\gev$) which does not contain a significant signal yield. 
The region $0.5 < \eleft < 1.0 \gev$ is defined to be the
\emph{neutral energy sideband}, while the region $\eleft<0.5\gev$
is referred to as the \emph{signal region}.

\subsection{Selection Efficiency}
\label{sec:Efficiency}
After applying all selection criteria we obtain a signal efficiency in
the Monte Carlo simulation of 
$(6.99 \pm 0.31) \times 10^{-4}$, where the error is purely from
the statistics of the signal MC. This number does not include
corrections for differences between data and simulation, which will
be described in Sec. \ref{sec:Systematics}.  
The number of background events expected for $\eleft<1.0\gev$, 
determined from simulation, is $269.1 \pm 9.6$, where the error 
is purely statistical.

\subsection{The Limit-Setting Procedure}
\label{sec:LimitSetting}

\begin{figure}[htb]
\begin{center}
\subfigure{%
\includegraphics[width=0.3\textwidth]{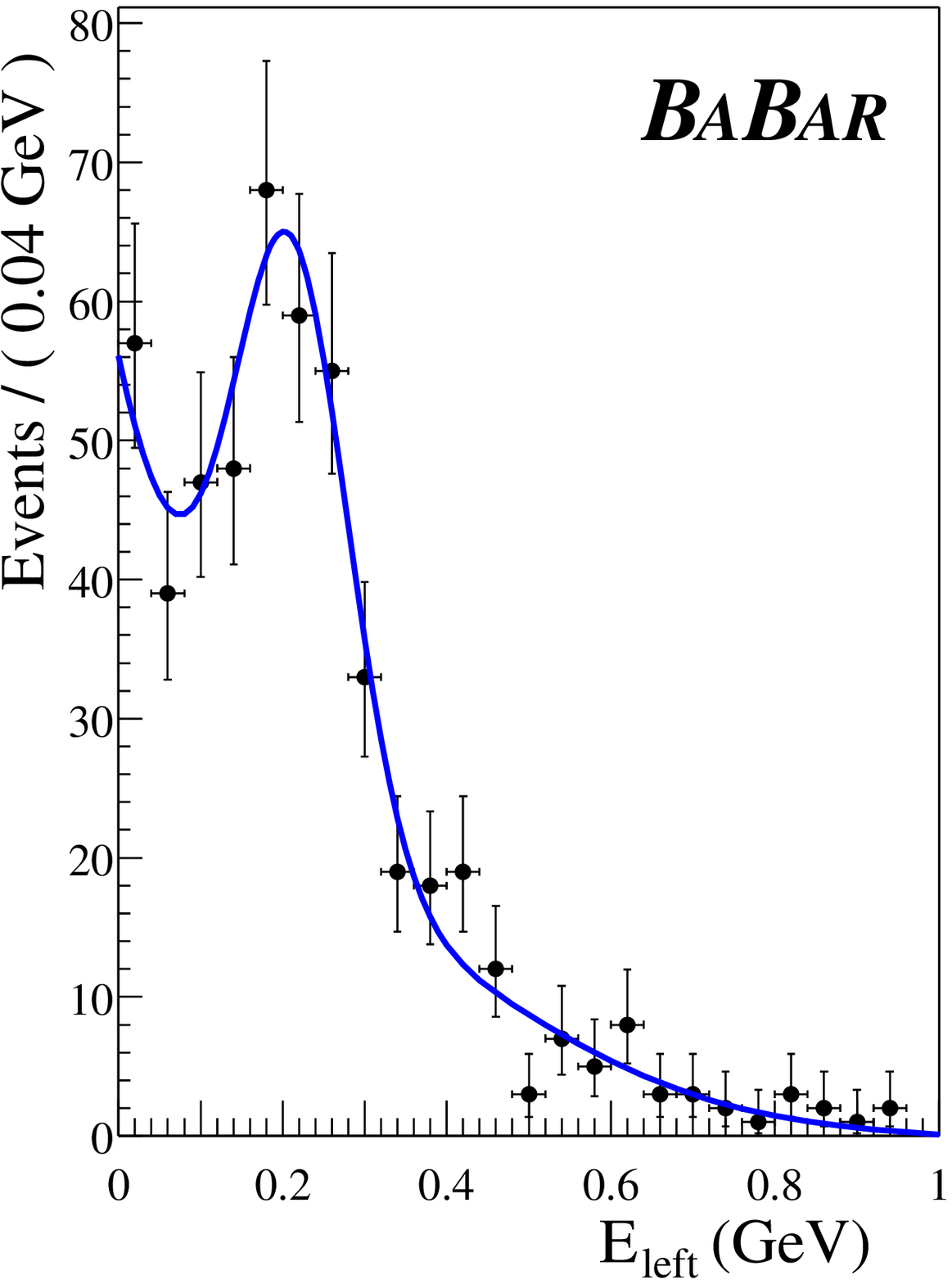}%
}
\subfigure{%
\includegraphics[width=0.3\textwidth]{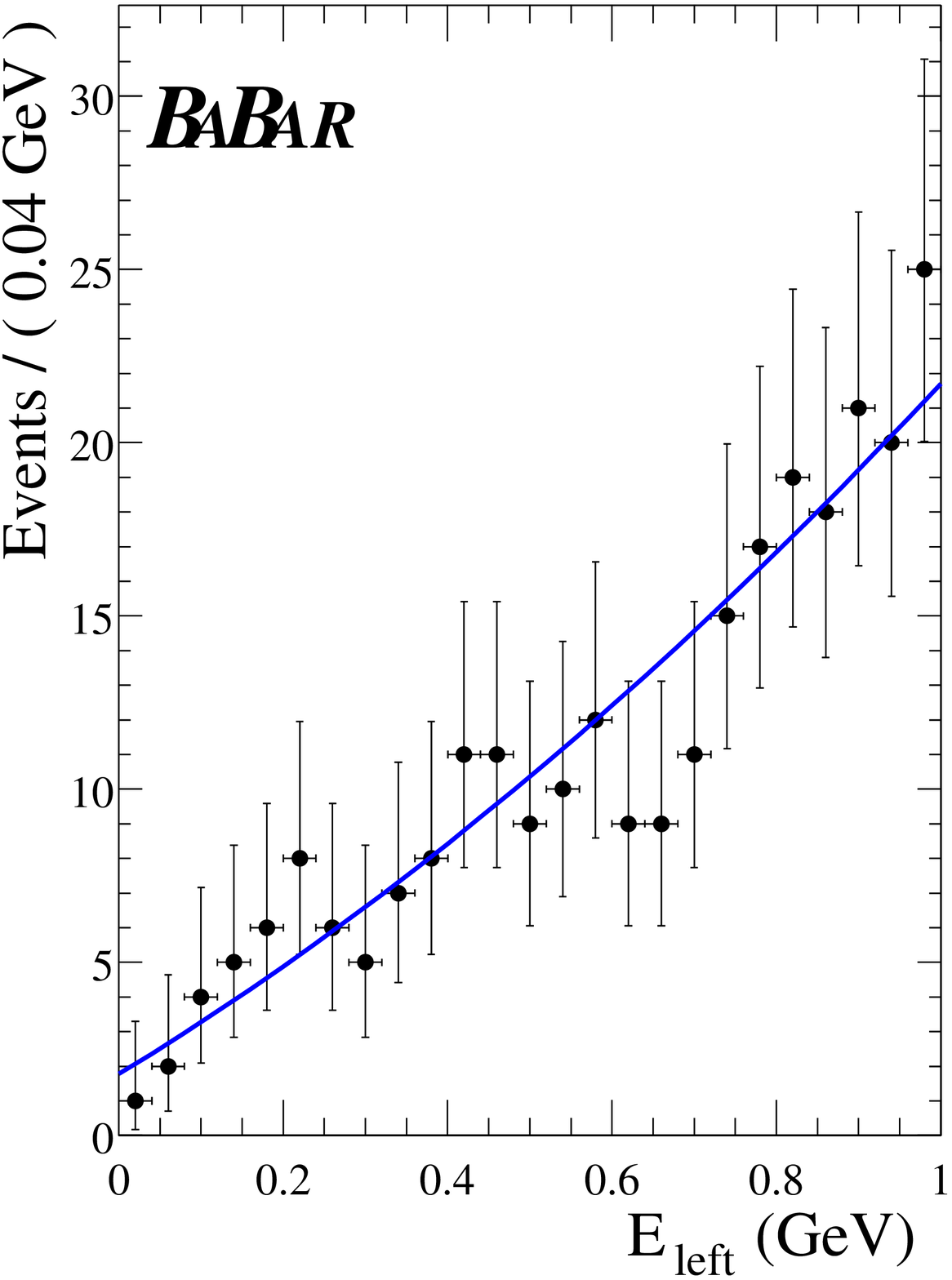}%
}
\end{center}
\caption{\label{fig:pdfs}%
The signal PDF (left) fitted to \eleft\ from the 
signal MC and the background PDF (right)
fitted to \eleft\ from the background MC. All selection criteria are
applied to the events in signal and background MC. The normalization of
the signal MC is arbitrary and the normalization of the background MC 
is fixed to the integrated luminosity. 
}
\end{figure}

We fit \eleft\ using an extended maximum likelihood function;
from this fit we extract the contributions to the distribution from
signal and background events. The likelihood function is of the following form:
\begin{eqnarray}
\label{eqn:elh_function}
\mathcal{L} = e^{-(\mu_s + \mu_b)} \prod_{i=1}^{n} \left(\mu_s \, F_s({\eleft}_i) + \mu_b \, F_b({\eleft}_i) \right)
\end{eqnarray}
where $\mu_s$ is the number of signal events, $\mu_b$ is the 
number of background events (both of which are obtained by fitting 
the selected data sample) and $n$ is
the total number of events in the selected sample.
The functions $F_s({\eleft})$ and $F_b({\eleft})$ are 
the Probability Density Functions (PDFs) which are obtained 
from the $\eleft$ distributions in signal and background MC.
The \emph{nominal} models for signal and background are shown
in Fig. \ref{fig:pdfs}. 

The \eleft\ distribution in the signal MC has several salient features.
The peak at $200\mev$ arises from the unassigned photon or 
\piz\ daughter of a \Dstarz\
on the semi-leptonic side. The peak at zero energy 
arises from events without remaining neutral energy, and above 
zero the spectrum falls exponentially under the peak from the 
\Dstarz\ neutrals, the result of neutrals in the MC arising from
beam background and hadronic split-offs.

The background PDF is a second-order polynomial rising 
monotonically between $0\gev$ and $1.0\gev$.
An estimate of the background from data is determined by extrapolating the
data \eleft\ sideband into the signal region using the background PDF. 
The result is a background estimate of $273.6 \pm 19.2$ events.

\begin{figure}[t]
\begin{center}
\subfigure{%
  \includegraphics[width=7.0cm,height=6.0cm]{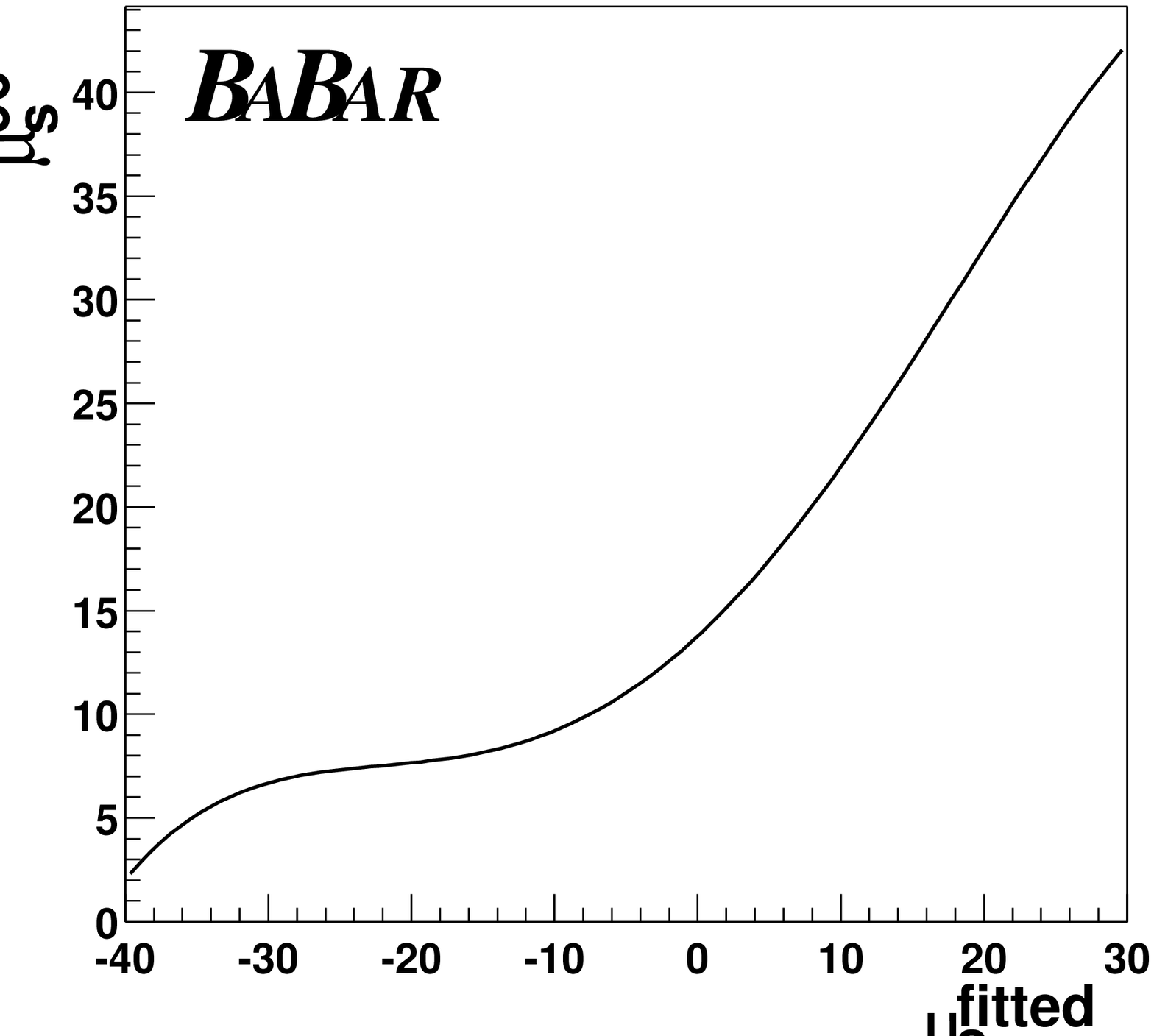}%
}
\hspace{0.5cm}
\subfigure{
  \includegraphics[width=7.0cm,height=6.0cm]{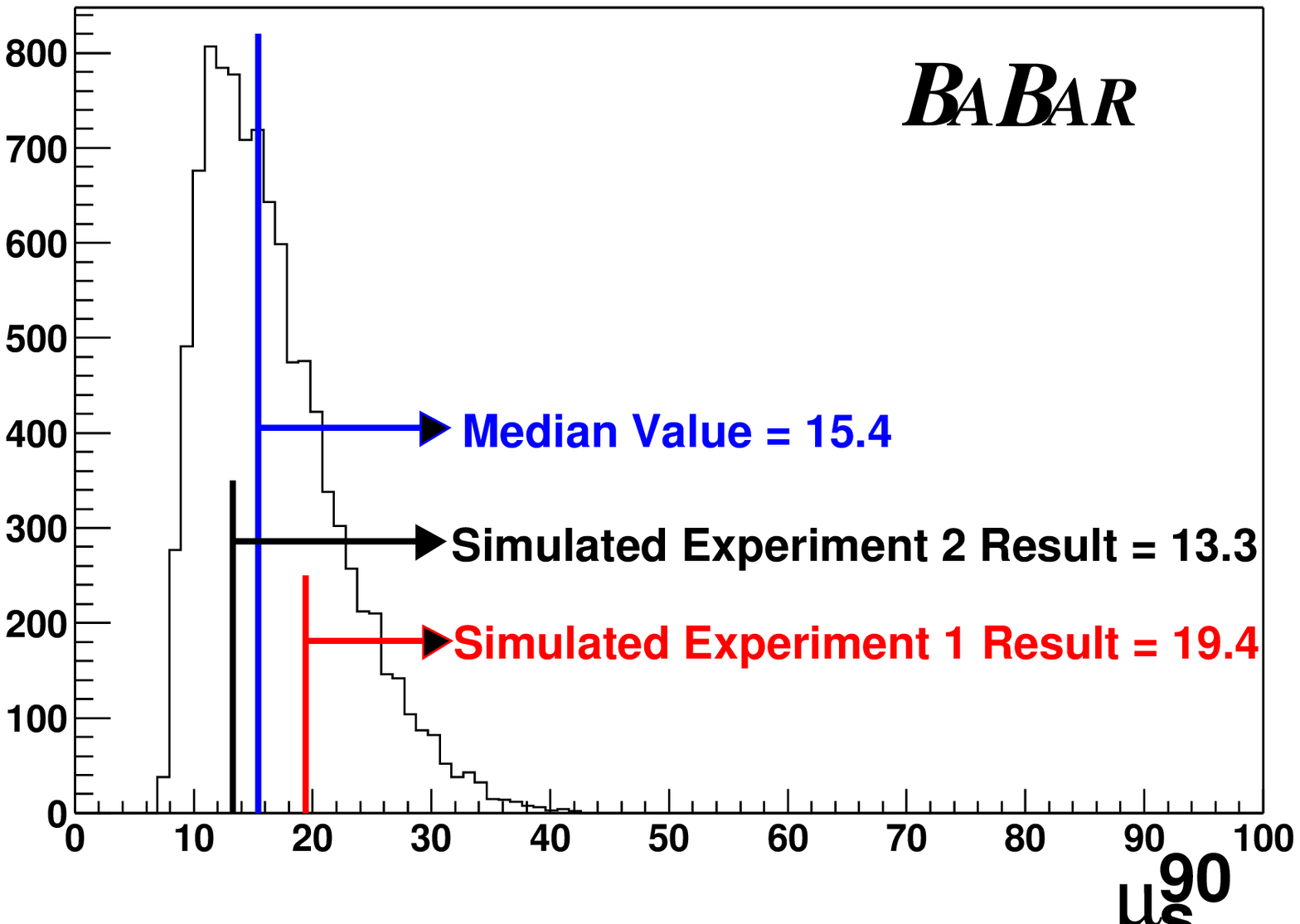}%
}
\end{center}
\caption{\label{fig:nominal_mus90}%
The relationship between $\mu_s^{90}$ and $\mu_s^{fitted}$
is shown in the left figure. In the right figure
is the distribution of $\mu_s^{90}$ obtained from fitting
10,000 toy MC experiments generated using the null signal
hypothesis. The black line indicates the median of this
distribution, which is taken as the nominal sensitivity
of the analysis  ($\mu_s^{90}=15.4$). The red and blue lines indicate
the results of applying the analysis
to two independent, background-only GEANT4 MC experiments.
}
\end{figure}

To determine the 90\% confidence limit, denoted $\mu_s^{90}$, for the measured
value of $\mu_s$ we use a modified frequentist 
approach based on that used by the LEP experiments in the 
Higgs search \cite{bib:higgs_cl}. This approach constructs an
estimator based on the fitted signal yield, 
$\mu_s^{fitted}$. In all following discussions, \emph{toy MC} refers to
\eleft\ spectra obtained by
randomly sampling the signal and background \eleft\ PDFs and treating the 
samples as a data set. The limit-setting method is outlined as follows:

\begin{enumerate}

\item Toy MC experiments are generated with our expected number of 
background events and a signal hypothesis, 
$\mu_{s}^{input} = \mu_{s}^{expected}$. 
Each toy experiment is fitted for the signal yield, $\mu_{s}'$. 
The collection of $\mu_{s}'$ forms a distribution. We
denote the normalized fitted signal distribution
for the original signal hypothesis as
$D_{s+b}(\mu_{s}',\mu_{s}^{input})$.

\item Toy MC is also generated with a signal hypothesis 
of zero events and with the expected number of background events. 
Again, each data sample is fitted and we obtain
the distribution of the fitted signal yield. We
denote the normalized distribution of the fitted signal
yield in the null signal hypothesis as $D_{b}(\mu_{s}')$.

\item For a specific data sample, a fit is performed  
using the likelihood function to obtain the number of 
signal events in the sample, $\mu_{s}^{fitted}$. The confidence
level (CL) for this measurement, for a signal hypothesis 
$\mu_{s}^{expected}$, is given by:
\begin{equation}
\label{eqn:mus10}
CL(\mu_{s}^{fitted},\mu_{s}^{expected}) \equiv  %
\frac{%
\int_{-\infty}^{\mu_{s}^{fitted}} D_{s+b}(\mu_{s}';\mu_{s}^{expected}) d\mu_{s}'%
}{%
\int_{-\infty}^{\mu_{s}^{fitted}}D_{b}(\mu_{s}') d\mu_{s}'%
}
\end{equation}

\item The 90\% confidence limit corresponding to this fitted 
signal yield, $\mu_{s}^{fitted}$, is defined by the requirement:
\begin{eqnarray}
CL(\mu_{s}^{fitted},\mu_{s}^{90}) & = & 1 - 0.90
\end{eqnarray}
A scan is performed over signal hypotheses until the minimum 
signal hypothesis which satisfies the requirement is determined. 
This signal hypothesis is $\mu_{s}^{90}$. The relationship between
$\mu_{s}^{90}$ and $\mu_s^{fitted}$ is given in Fig. \ref{fig:nominal_mus90}.

\end{enumerate}

We use the background-only hypothesis to determine the nominal 
sensitivity for the analysis. The nominal background PDF is
used to generate 10,000 toy MC experiments, each with
a total number of events Poisson-distributed around the
nominal background expectation of $269 \pm 9.6$ events. 
The \eleft\ distribution
so generated for each experiment is fitted with the likelihood
function, and the curve in Fig. \ref{fig:nominal_mus90} is used to
obtain the corresponding $\mu_{s}^{90}$. 
The distribution of $\mu_s^{90}$ for these toy experiments
is shown in Fig. \ref{fig:nominal_mus90}, and the median of this
distribution provides our nominal sensitivity, $\mu_s^{90} = 15.4$.

A test of the method is performed using two independent samples
of GEANT4 MC, each with a total luminosity equal to that
of the data set. Each simulated experiment contains
only background events and has all analysis selection criteria applied. 
The results obtained from fitting these two
experiments, $\mu_s^{90} = 13.3$ and $\mu_s^{90} = 19.4$ are 
in good agreement with the nominal sensitivity and are illustrated in Fig.
\ref{fig:nominal_mus90}. The results of these
GEANT4 MC experiments are used in 
Sec. \ref{sec:Systematics_CL} to explore 
potential systematic effects on the limit-setting procedure.

The strength of this method is that it takes full advantage of
the information contained in the shape of the signal and background and
is insensitive to the absolute background normalization. 
It relies on the accuracy of the simulation,
in particular the simulation of the calorimeter for low
neutral cluster energy. The systematic effects on the limit-setting
procedure, as well as on the signal efficiency, are discussed in
the next section.

\section{Evaluation of Systematic Uncertainties}
\label{sec:Systematics}
In this section we discuss the systematic uncertainties affecting the
limit-setting procedure and the calculation of the efficiency.

\subsection{Systematic Effects for Limit-Setting}
\label{sec:Systematics_CL}

Table \ref{tab:cl_systematics} lists the sources of systematic
uncertainty and their effects on the overall expected sensitivity
of a pair of independent ``experiments'' derived from the
background MC sample. Since MC is used to
generate the PDFs for signal and background, effects of concern are
those which might cause the true background or signal
distributions to differ from the simulation.
For each source of potential systematic uncertainty listed below,
a CL curve is calculated (as in Fig. \ref{fig:nominal_mus90}) by
generating toy MC with a systematically modified model and 
fitting with the nominal model.

To explore the potential differences between data and MC
when validating the background PDF, several control samples have been developed
by performing the semi-leptonic reconstruction 
on data and MC and applying selection
criteria to the remainder of the event:

\begin{itemize}
\item {\bf ($\btdlnux$) + (anything)}: The \btdlnux\ reconstruction 
algorithm is applied and no restrictions are placed on the signal side. 

\item {\bf ($\Bz \to \Dstarp\ell^{-}\nu$) + (1 track) }: The semi-leptonic
reconstruction
algorithm is applied, but semi-leptonic candidates are kept where the best
candidate is reconstructed as a $\Dstarp\ell^{-}\nu$. We
then require only one track remaining in the event. This sample
has a topology similar to the signal sample.

\item {\bf ($\btdlnux$) + (2 tracks)}: The \btdlnux\ reconstruction algorithm
is applied and the signal side is required to contain exactly
two charged tracks. This requirement enforces a mis-reconstructed 
signal side, since total event charge is non-zero. 
\end{itemize}

The difference between data and simulation for each of these
samples is used to construct a correction function for the PDFs.
The PDFs are multiplied by the correction function and re-normalized.
The upper limit is recalculated using the corrected PDFs to generate
the toy MC samples discussed in Sec. \ref{sec:Analysis}, 
while fitting is still performed with the nominal model. The results
of this study are given in Table \ref{tab:cl_systematics}.

Motivated by the data/MC yield disagreement in the control samples, 
we explored the effects of a significant difference between the
actual background and the background expectation. Given the expected
background from MC ($269.1 \pm 9.6$ events), we varied the
input background hypothesis to our toy MC by $\pm10$ and $\pm20$ events and used the 
nominal likelihood function to generate CL curves for each case.
The result was a change in the nominal sensitivity of $\mu_s^{90}$ of
no more than $0.5$ events (Table \ref{tab:cl_systematics}). 
Therefore we concluded that if the actual background
in data is different than the expectation from MC, the resulting 
change in the limit is small. 

The choice of background parametrization has also been
explored as a possible effect on the limit-setting procedure. 
Different models (second-order polynomial, Gaussian tail, KEYS \cite{html:keys},
histogram) were used to represent the background \eleft\ 
spectrum and then generate toy MC. These toy samples
were again fitted with the nominal model and the variation in the
confidence limit curve was explored. The results are given in
Table \ref{tab:cl_systematics}. Limit curves calculated from this
variation in the background model yielded a lower limit on the
signal yield, suggesting that the second-order polynomial leads to
a conservative limit.

A feature at or near zero energy in the \eleft\ spectrum, unmodelled in the
Monte Carlo, is another possible effect in the background that
could fake a signal. Such a feature is suggested by studying the 
$\Bz \to \Dstarp\ell^{-}\nu$ + 1 track control sample, 
which  suggested a maximal difference near
zero energy in data and MC of a factor of two. To explore the effect of such a 
feature, toy MC is generated with the background PDF enhanced 
by a factor of two near zero. This toy MC is then fit using the nominal
signal and background models to study the variation in the confidence 
limit. The results are given in Table \ref{tab:cl_systematics}.

To validate the signal PDF, a control sample of 
$\btdlnux$,$\Bu \to \Dstarzb\ell^{+}\nu_{\ell}$ is used to 
check the difference between data and simulation
after making all semi-leptonic and signal-side requirements. This sample has
a semi-leptonic side as in the signal analysis, and a signal side which is
also reconstructed as an independent \bbtdlnux. We then construct
a subsample of these events where a photon or \piz\ could be associated with
the \Dzb\ on the signal side to make a \Dstarzb. Having attempted to
account for the neutral from \Dstarzb\ decay on the signal side, the
neutral energy spectrum should resemble that for signal
events. However, this subsample has more neutral energy present from
hadronic splitoffs, and the correct photon or \piz\ is not always associated
with the \Dzb\ to make the \Dstarzb. Therefore, a comparison is made
between data and the corresponding MC sample and not between data and signal MC. 
The sample suggests that there may be a difference of $\sim 20\mev$
in the overall position of the peak near $200\mev$. The effects
of such a difference are given in Table \ref{tab:cl_systematics}.

\begin{table}[htb]
\caption{\label{tab:cl_systematics}%
Sources of potential systematic uncertainty have been explored and their
effects on the expectation for $\mu_s^{90}$ observed. The
simulated experiments come from two independent samples of 
full MC, each with luminosity equivalent to the data
luminosity. Despite
the many potential differences between data and simulation, the
variation in the upper limit is small copared to the variation
of limit expected for the zero signal hypothesis. 
}
\begin{center}
\begin{tabular}{|l|c|c|c|}
\hline
Systematic Variation  & Control Sample used& $\mu_s^{90}$  & $\mu_s^{90}$\\
                      & to guide the variation & Simulated  & Simulated \\
                      &                        & Experiment 1 & Experiment 2 \\
\hline
None (Nominal Model)  & None           &   19.4            & 13.3            \\
\hline
Data/Simulation       & $\btdlnux$ + anything                  & 18.0 & 12.4 \\
Shape Difference      & $\Bz \to \Dstarp\ell^{-}\nu$ + 1 track & 18.6 & 13.0 \\
                      & $\btdlnux$ + 2 tracks                  & 17.8 & 12.2 \\
\hline
Incorrect Background Estimate & & &\\
-20 Events                    & & 19.0  & 12.0 \\
-10 Events                    & All & 19.3 & 13.1\\
+10 Events                    &  & 19.7 & 13.6 \\
+20 Events                    &  & 19.8  & 13.7 \\
\hline
Gaussian Tail Background Model   & None   &  17.1             & 12.0            \\
KEYS Background Model       & None   &  17.4             & 12.0            \\
Histogram Background Model  & None   & 17.2            & 11.8            \\
\hline
Factor of 2           & & & \\
enhancement in        & $\Bz \to \Dstarp\ell^{-}\nu$ + 1 track & 16.5 & 11.5 \\
background near       & & & \\
$\eleft = 0.0\gev$    & & & \\
\hline
$+20\mev$ Shift in    &                  & & \\
Signal  PDF Gaussian  & $\btdlnux$,$\Bu \to \Dstarzb\ell^{+}\nu_{\ell}$ & 20.2 & 13.9 \\
\hline
$-20\mev$ Shift in    &                  & & \\
Signal  PDF Gaussian  & $\btdlnux$,$\Bu \to \Dstarzb\ell^{+}\nu_{\ell}$ & 18.9 & 13.0 \\
\hline
\end{tabular}
\end{center}
\end{table}

In almost all cases that were investigated, the nominal model yields the most conservative
CL curve. Additionally, the variation in the upper limit expectation is small
compared to the overall variation of the limit for the zero signal
hypothesis. We therefore use the nominal model to set the limit,
having determined that the systematic effects mentioned above
have negligible impact on the result.

\subsection{Systematic Effects for Signal Efficiency}
\label{sec:Systematics_efficiency}

The efficiency calculated in Sec. \ref{sec:Efficiency} needs to be corrected for
differences in data and MC. The effects of several systematic
uncertainty have been evaluated. The results are expressed in a
correction factor and a systematic error associated with the
correction. The most prominent sources of systematic uncertainties
are listed in Table \ref{tab:systematics}.

\begin{table}[bt]
\caption{\label{tab:systematics}%
The corrections which are applied in this analysis on the signal
efficiency, as well as the error associated with each correction.%
}
\begin{center}
\begin{tabular}{|l|c|c|} 
\hline
Source              & Correction Factor &  Correction Error \\ 
\hline\hline                                                
B-Counting          & 1.0               &  0.011            \\
\hline                                                      
Semi-Leptonic       &                   &                    \\
Reconstruction Efficiency  & 0.958             &  0.051            \\
\hline                                                      
Neutral  Energy     & 1.0               &  0.025            \\
\hline                                                      
Tracking Efficiency &                   &                   \\
(signal track only) & 1.0               &  0.008            \\
\hline                                                      
Particle ID         &                   &                   \\
(signal track only) & 0.836             &  0.042            \\
\hline                                                      
\hline                                                      
Total               & 0.801             &  0.063            \\
\hline

\end{tabular}
\end{center}
\end{table}

The semi-leptonic reconstruction efficiency correction is evaluated by using
a sample of $N_2$ double-reconstruction events (\btdlnux, \bbtdlnux) and a sample of
$N_1$ single-reconstruction events in the same data set, 
where the single-reconstruction sample excludes
double-reconstructed events. To isolate the effects of the semi-leptonic
reconstruction, we remove the event preselection requirements
which are unrelated to
the selection of the semi-leptonic decay. The ratio of the sizes of
these two samples is given by:
\begin{eqnarray}\label{eqn:doubletag}
\frac{N_1}{N_2} & = & \frac{2(1-\varepsilon)}{\varepsilon}
\end{eqnarray}
where $\varepsilon$ is the probability of selecting a second $B$ meson
in the mode \btdlnux\ having already selected one $B$ meson in
that mode. %
Equation \ref{eqn:doubletag} is solved for $\varepsilon$, and
the ratio of $\varepsilon$ in data and simulation is taken as 
a correction on the efficiency,
\begin{eqnarray}
\frac{\varepsilon_{data}}{\varepsilon_{MC}} & = & 0.958 \pm 0.051 \,.
\end{eqnarray}
The correction of the signal efficiency due to particle identification (PID)
is performed as follows. Using control samples of electrons and 
muons from data, the performance of the kaon veto and lepton
identification is evaluated as a function of lab
momentum ($p$) and polar angle ($\theta$). This performance
translates into a weight for an identified track with a given momentum. 
Integrating this weight over the momentum spectrum of the signal
MC for true one-prong leptonic \taup\ decays provides
the efficiency of the selection in data.

The efficiency in the MC is evaluated 
by comparing the true identity of a \taup\ daughter to
the reconstructed identity. The ratio of the data efficiency
to the MC efficiency is taken as the signal efficiency
correction from particle identification. The statistical uncertainties 
on the signal MC sample size and PID weights translate into a systematic
error on the correction. This error is dominated by the signal MC
sample size.

We apply the total correction listed in Table \ref{tab:systematics} 
to the signal 
efficiency from MC to obtain the signal efficiency,
$(5.60 \pm 0.25 (stat.) \pm 0.44 (syst.)) \times 10^{-4}$. 
The uncertainty on each correction 
is propagated through to a systematic error on the 
signal efficiency, while the statistical uncertainty 
derives from the signal MC sample size.

Systematic uncertainty is incorporated into the branching fraction limit
as follows: for each branching fraction hypothesis, the corresponding signal
hypothesis is determined by sampling 10,000 times from a Gaussian distribution
about the mean of the signal efficiency, where the width of the
Gaussian is given by the systematic error on the signal efficiency. 
From each signal hypothesis a toy MC experiment is generated and fitted
using the likelihood function. This is repeated for many branching
fraction hypotheses and the CL curve, including systematic error, is 
calculated.

\section{Results}
\label{sec:Physics}

\begin{figure}[htb]
\begin{center}
\includegraphics[width=0.45\textwidth]{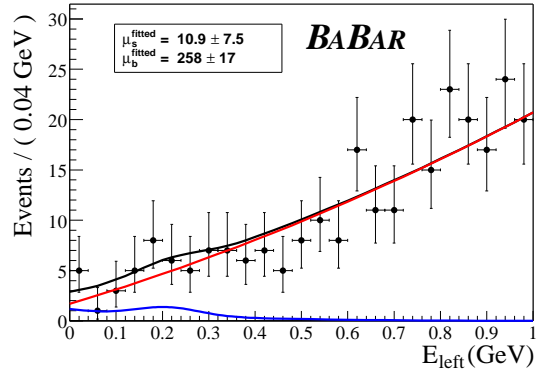}%
\end{center}
\caption{\label{fig:fit_result}%
The \eleft\ spectrum in data from zero to $1.0\gev$ and the
fit of the likelihood function to the spectrum. 
}
\end{figure}

We apply all selection criteria in the analysis and fit the \eleft\ spectrum in the 
data below $1.0\gev$ using the nominal likelihood function. 
The result of the fit is given in Table \ref{tab:fit_result}
and shown in Fig. \ref{fig:fit_result}. 

\begin{table}[htb]
\caption{\label{tab:fit_result}%
The results of the fit of the likelihood function to data is
given below.
}
\begin{center}
\begin{tabular}{|c|c|}
\hline
Fit Parameter     &     Fitted Value      \\
\hline\hline
$\mu_{s}^{fitted}$&     $10.9 \pm 7.5$    \\
\hline
$\mu_{b}^{fitted}$&     $258 \pm 17$      \\
\hline
\end{tabular}
\end{center}
\end{table}

The background yield of $258 \pm 17$ is in agreement with the 
expectation from extrapolating sideband data into the \eleft\
signal region, $274 \pm 19$. 

\begin{figure}[htb]
\begin{center}
\includegraphics[width=0.45\textwidth]{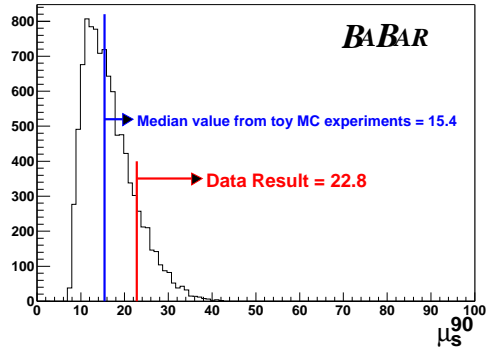}%
\end{center}
\caption{\label{fig:compare_data_mc}%
The nominal sensitivity (15.4), obtained from the median of 10,000 toy
MC experiments, is compared to the result from data (22.8). 
}
\end{figure}

We apply our confidence limit curve to the fitted number of signal events 
(Fig. \ref{fig:nominal_mus90}) and determine the 90\% confidence
limit on the signal yield to be
\begin{eqnarray}
\mu_s^{90} & = & 22.8,
\end{eqnarray}
which yields a limit on the branching fraction at the 90\%
confidence level of
\begin{eqnarray}
\mathcal{B}(\btn) & < &  4.6 \times 10^{-4} \, \textrm{at the 90\% CL}
\end{eqnarray} 
where we have assumed equal production of $\Bu\Bub$ and $\Bz\Bzb$ at the
\FourS\ resonance. The expectation from studies in MC containing
zero signal events was given in Sec. \ref{sec:Analysis} as 
$\mu_s^{90} = 15.4$; the comparison of the expected sensitivity and the
measured result is shown in Fig. \ref{fig:compare_data_mc}.
\begin{figure}[htb]
\begin{center}
\includegraphics[width=0.45\textwidth]{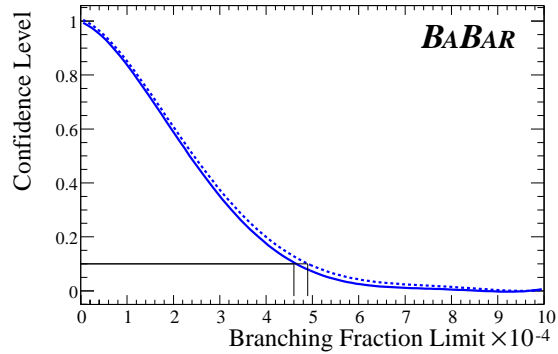}%
\end{center}
\caption{\label{fig:cl_vs_br}%
The relationship between the confidence level and the
branching fraction limit both in the case where systematic error
is (dashed) and is not (solid) incorporated. Lines are drawn
at the locations of the 90\% confidence limits for both cases.
}
\end{figure}

After incorporating the systematic uncertainty
on the signal efficiency (Fig. \ref{fig:cl_vs_br}) 
we find an upper limit on the branching fraction of
\begin{eqnarray}
\mathcal{B}(\btn) & < & 4.9 \times 10^{-4} \, \textrm{at the 90\% CL}
\end{eqnarray} 

A separate \babar\ analysis to search for \btn\ has also been performed where
$B$ mesons are resonstructed via the semi-exclusive hadronic modes 
$\Bub \to D^{(*)0} X^{-}$, where $X^{-}$ 
represents a combination of up to five charged
pions or kaons and up to two \piz\ candidates.
The semi-exclusive hadronic reconstruction analysis is 
statistically independent 
of the present work and has obtained the limit \cite{se_btn}
\begin{eqnarray}
\mathcal{B}(\btn)<7.7\times10^{-4} \, \textrm{at the 90\% CL}
\end{eqnarray}
The results of this statistically independent analysis are combined
with that reported here using the same modified frequentist
method. For each analysis, 5000 toy MC
experiments are generated each for a branching
fraction hypothesis between $1 \times 10^{-4}$ and $10 \times 10^{-4}$
in steps of $2.5 \times 10^{-6}$. The number of events in each toy
experiment that contains signal and background is denoted $n_{(s+b)}$. 
For each branching fraction hypothesis a separate sample 
of 5000 background-only toy experiments is also generated;
the number of events in a given experiment in this sample is denoted
$n_{(b)}$. For a given hypothesis $i$ (where $i=(s+b)$, $(b)$, or $data$)
the likelihood estimators, $Q_{i}$, are calculated in each toy experiment:
\begin{eqnarray}
Q_{i} & = & \frac{\mathcal{L}(n_{i} | s+b)}{\mathcal{L}(n_{i} | b)}
\end{eqnarray}
where $\mathcal{L}(n_{i} | s+b)$ is the likelihood of a given
toy experiment with $n_{i}$ events being consistent with 
the $(s+b)$ hypothesis (in equation
\ref{eqn:elh_function}, $\mu_s$ is set to the
number of signal $s$, corresponding to the branching fraction hypothesis,
and $\mu_b$ is set to the number of background $b$). 
Systematic errors are incorporated by
each analysis into the $Q_{i}$. They are combined as follows:
\begin{eqnarray}
Q_{i}^{\textrm{combined}} & = & Q_{i}^{\textrm{semi-leptonic}} \times Q_{i}^{\textrm{hadronic}}.
\end{eqnarray}
The confidence level for a particular branching ratio hypothesis is given by
counting the number of experiments, $N_{(Q_i<Q_{data})}$, 
that have a value of $Q_{(s+b)}$ ($Q_{(b)}$) less than that for 
data ($Q_{data}$) and taking the ratio
\begin{eqnarray}
CL_{s} & = & \frac{ N_{({Q_{(s+b)}<Q_{data}})} }{ N_{({Q_{(b)}<Q_{data}})} }.
\end{eqnarray}
From this a combined \babar\ result is obtained:
\begin{eqnarray}
\mathcal{B}(\btn)_{combined}<4.1\times10^{-4} \, \textrm{at the 90\% CL}.
\end{eqnarray}

\section{Summary}
\label{sec:Summary}
We have performed a search for the decay 
process \btn. To accomplish this a sample of
semi-leptonic $B$ decays (\btdlnux) has been used to
reconstruct one of the $B$ mesons and the remaining information in
the event is searched for evidence of \btn. We find no evidence for this
decay process and set a preliminary limit on its branching fraction of
\begin{eqnarray*}
\mathcal{B}(\btn) & < & 4.9 \times 10^{-4} \, \textrm{at the 90\% CL}
\end{eqnarray*}
By combining this analysis with a statistically independent
\btn\ search performed using a semi-exclusive $B$ reconstruction 
we find the preliminary combined limit:

\begin{eqnarray*}
\mathcal{B}(\btn)_{combined}<4.1\times10^{-4} \, \textrm{at the 90\% CL}
\end{eqnarray*}

\section{Acknowledgments}
\label{sec:Acknowledgments}


\input{acknowledgements}

\end{document}

%% file: user_symbols.tex
\def\btn {\ensuremath{B^{+} \to \tau^{+} \nu_{\tau}}}
\def\btdlnux {\ensuremath{\Bub \to \Dz \ell^{-} \bar{\nu}_{\ell} X}}

\def\bbtdlnux {\ensuremath{\Bu \to \Dzb \ell^{+} \nu_{\ell} X}}
\def\Dl {\ensuremath{D\ell}}
\def\cosbydl {\ensuremath{\cos{\theta_{B,\Dl}}}}
\def\eleft {\ensuremath{E_{\mathrm{left}}}}

%% file: authors_win2003.tex
\begin{center}
\small

The \babar\ Collaboration,
\bigskip

%
B.~Aubert,
R.~Barate,
D.~Boutigny,
J.-M.~Gaillard,
A.~Hicheur,
Y.~Karyotakis,
J.~P.~Lees,
P.~Robbe,
V.~Tisserand,
A.~Zghiche
\inst{Laboratoire de Physique des Particules, F-74941 Annecy-le-Vieux, France }
% ...

\end{center}\newpage

%% file: acknowledgements.tex
We are grateful for the 
extraordinary contributions of our \pep2\ colleagues in
achieving the excellent luminosity and machine conditions
that have made this work possible.
The success of this project also relies critically on the 
expertise and dedication of the computing organizations that 
support \babar.
The collaborating institutions wish to thank 
SLAC for its support and the kind hospitality extended to them. 
This work is supported by the
US Department of Energy
and National Science Foundation, the
Natural Sciences and Engineering Research Council (Canada),
Institute of High Energy Physics (China), the
Commissariat \`a l'Energie Atomique and
Institut National de Physique Nucl\'eaire et de Physique des Particules
(France), the
Bundesministerium f\"ur Bildung und Forschung and
Deutsche Forschungsgemeinschaft
(Germany), the
Istituto Nazionale di Fisica Nucleare (Italy),
the Foundation for Fundamental Research on Matter (The Netherlands),
the Research Council of Norway, the
Ministry of Science and Technology of the Russian Federation, and the
Particle Physics and Astronomy Research Council (United Kingdom). 
Individuals have received support from 
the A. P. Sloan Foundation, 
the Research Corporation,
and the Alexander von Humboldt Foundation.